\documentclass[fleqn,11pt]{wlscirep}
\usepackage[utf8]{inputenc}
\usepackage[bottom]{footmisc}
\usepackage{wrapfig}
\usepackage{float}
\usepackage{authblk}
\usepackage{tabto}
\usepackage{hyperref}
\usepackage[T1]{fontenc}
\title{EMRs with Blockchain : A distributed democratised Electronic Medical Record sharing platform}

\author[1]{\rule{1in}{0pt}Sanket Shevkar} 
\author[2]{Parthit Patel}
\author[3,*]{Saptarshi Majumder }
\author[4]{\\  \rule{1in}{0pt} Harshita Singh}
\author[5]{Kshitijaa Jaglan}
\author[6,*]{Hrithwik Shalu}

\affil[1]{Symbiosis Institute of Technology, Pune}
\affil[2]{Savitribai Phule Pune University, Pune}
\affil[3]{Indian Institute of Technology, Bombay}

\affil[4]{Vivekanand Education Society's Institute of Technology, Mumbai}
\affil[5]{International Institute of Information Technology, Hyderabad}
\affil[6]{Indian Institute of Technology, Madras}
\affil[*]{Corresponding authors: majumder.saptarshi@iitb.ac.in, ae18b116@smail.iitm.ac.in}


\begin{abstract}

Medical data sharing needs to be done with the utmost respect for privacy and security. It contains intimate data of the patient and any access to it must be highly regulated. With the emergence of vertical solutions in healthcare institutions, interoperability across organisations has been hindered. The authors of this paper propose a blockchain based medical-data sharing solution, utilising Hyperledger Fabric to regulate access to medical data, and using the InterPlanatory File System for its storage. We believe that the combination of these two distributed solutions can enable patients to access their medical records across healthcare institutions while ensuring non-repudiation, immutability and providing data-ownership. It would enable healthcare practitioners to access all previous medical records in a single location, empowering them with the data required for the effective diagnosis and treatment of patients. Making it safe and straightforward, it would also enable patients to share medical data with research institutions, leading to the creation of reliable data sets, laying the groundwork required for the creation of personalised medicine.

\end{abstract}
\begin{document}

\flushbottom
\maketitle


\section{Introduction}

\tab With the rapid developments in the healthcare industry, the amount of electronic data generated by medical institutions has grown tremendously. Medical data despite being highly personal, is not controlled or managed by the patients themselves. Throughout the industry, the authority to manage data lies with the data producer. If a patient wishes to access their own personal data, or share it with another healthcare professional, they require approval from the organisation managing their medical data. Thus, patients in effect, do not own or control their own medical records. Medical data contains sensitive information about an individual. Sharing such data requires a secure platform, demanding high security and privacy along with strict access control restrictions. The lack of industry wide standards for sharing medical data has given rise to data silos. It has led to vertical development of data management solutions within organisations, stifling interoperability. This also leads to inaccessibility and provider lock-in. Hence, there is a need for a widely accessible solution, which would provide interoperability in a heterogeneous environment and provide the requisite security and privacy that medical records demand. \\
\newline
\tab Blockchain technology is poised as a potent solution for the medical industry. Touted as one of the most substantial technologies of the future, it originated primarily as an undergird to decentralising long standing notions of banking. Besides enabling the rise of cryptocurrencies, blockchain technology has huge implications for data sharing and accountability across industries. All transactions on the blockchain network are written to a cryptographically secure immutable ledger. The ledger, based on a distributed peer-to-peer network, enables the development of solutions requiring high accountability and transparency. Blockchain technology also enables data sovereignty, embedding data ownership as an important characteristic. This allows development of solutions that enable users to manage their data with a desired granularity of access and modification permissions. Modification or access to patient data can be routed through the blockchain network. This would result in a transaction being produced anytime access to patient records are requested. These transactions are written to the immutable ledger in an append only fashion. Ensuring that any change made to the patient data is traceable, it would make medical data resistant to fraud and fabrication. It also allows for the development of horizontal, industry wide solutions. It can act as a secure access point available to healthcare institutions, encapsulating implementation details and promoting interoperability.\\
\newline
\tab The authors of this paper propose a data sharing system for Electronic Health Record (EHR) sharing using the Hyperledger Fabric platform. Hyperledger Fabric is a private permissioned blockchain solution. Users would have to register themselves to an authority in the blockchain network to access it. Hyperledger fabric uses channels to create a private communication medium between two or more entities on the blockchain network, thus ensuring that data shared between participants of the channel is not accessible to other members of the network. The project also incorporates the usage of Interplanetary File System (IPFS) for data storage. IPFS presents itself as a powerful permanent data storage solution accessible across heterogeneous systems. Through a distributed peer-to-peer file storage system, it provides self distribution, removing dependencies on a content distributor. Pairing IPFS with blockchain technology presents a compelling solution to creating an immutable distributed digital solution for heterogeneous systems. The solution aspires to give patients control of their own medical data, making health data accessible across the globe, while the underlying blockchain technology ensures the privacy and integrity of the data stored.

\section{Related Work}

Relevant work has been done in the field exploring the benefits of blockchain technology in the healthcare domain. Many viable solutions have been presented. In this section, we discuss the relevant work done.

The authors\cite{leeming2019ledger} of the paper explore Personal Health Records as a potential means of providing patients fine-grained, personalised and secure access to their medical records. Utilising blockchain and distributed ledger technology, the authors of the paper propose 'Ledger of Me', a PHR solution that puts the patients in charge of their own data.

MediBloc\cite{team2018medibloc},  a commercial product uses the Qtum public blockchain. Real data is stored on a Distributed Hash Table on the IPFS, with the meta-data stored on the blockchain . Access control is dictated by smart contracts and has a transaction fee associated. 

In the paper, the authors\cite{thwin2018blockchain} discuss a model for a blockchain based personal medical record sharing model. They expound on an append-only model, the attacks it is vulnerable to and a solution to overcome it. 

The Authors\cite{nizamuddin2018ipfs} discuss a viable solution using the Ethereum blockchain and the IPFS network for verifying the authenticity of online content, specifically online-books. It presents an extensible model that can be used to extend its functionality beyond online-books to other forms of digital content.

\section{Background}
\subsection{Blockchain}
Blockchain technology, introduced through the Bitcoin blockchain \cite{nakamoto2019bitcoin}, is a decentralized, distributed ledger of transactions that maintains verified transactions. A genesis block, is the first block of the blockchain. This block has no transactions. A transaction on the blockchain, is the primary building block of the system. Using the transactions that are created on the blockchain network, subsequent blocks are generated through a consensus mechanism, after being verified by special nodes on the network called miners. These blocks of transactions that have been approved by the miners, are appended to the blockchain. Nodes keep an updated copy of the blockchain network, as blocks are appended to the original chain. Transactions are generated whenever there is a transfer of cryptocurrency or input data between participants of the network, or through the execution of a smart contract. Transactions that are pending to be verified are pulled by the miner nodes. They are verified through the consensus mechanism, and a new block is created. This block is added to the main chain in an append only fashion,  thus creating an immutable chronologically ordered ledger. 

\subsection{Consensus Mechanisms}
The consensus mechanism of a blockchain determines the protocol for selection of the block to be added to the chain. It is used for verifying the transactions and the validity of the block contents. Changing the parameters of the consensus mechanism can allow for the modification of block characteristics. It is a mechanism for creating agreement among the participants of the blockchain network about the world state of the blockchain. It contains the rules for transaction validation, block addition and addresses partition when the network is forked or partitioned. Bitcoin and Ethereum blockchain\cite{buterin2014ethereum} networks employ a Proof of Work (PoW) algorithm. It requires miners to invest huge computational power solving a cryptographic puzzle for the addition of a new block. Hyperledger Fabric\cite{androulaki2018hyperledger} does not employ a consensus algorithm, but consensus is achieved through an ordering service. It can use Raft, Kafka or Solo for the ordering mechanism. Invocation of a chaincode leads to the creation of a proposal. Special nodes called endorsers simulate the proposal by executing the chaincode installed on the blockchain. The endorsers respond with an endorsement response. Once the requisite number of endorsements are received, the transaction is sent to the ordering service, which creates the block and is responsible for the broadcast and delivery of the created blocks to the nodes with the blockchain ledger in the network.

\subsection{Hyperledger Fabric}
The Hyperledger Fabric platform \cite{androulaki2018hyperledger} is a private-permission blockchain solution. It is a platform designed for distributed ledger solutions with a modular architecture. It addresses the issues of confidentiality, flexibility and scalability that other blockchain solutions fail to meet. Unlike other widely-used blockchain solutions, such as Ethereum and Bitcoin which are public and permissionless, the Hyperledger Fabric platform only allows authorized individuals to make transactions on the network. It uses a Membership Service Provider to enroll members to the blockchain network. It is designed to support pluggable implementations of components catering to the requirements of different business solutions.

\subsubsection{Distributed Ledger}
A decentralised ledger is used to store all the transactions that take place on the blockchain network. This ledger is decentralised in nature as it is replicated across multiple participants of the blockchain network. The information recorded on the ledger is append-only. This ensures the immutability of the blockchain network, while also storing information about data provenance. The ledger in Hyperledger Fabric consists of two components, the world state and the transaction log. All the transactions that have resulted in the current state of the world state, are recorded in the transaction log. The ledger thus, is in effect, a combination of the transaction log and the world state database.

\subsubsection{Smart Contracts}
Smart Contracts are a key mechanism for ensuring the seamless execution of use-case specific logic in the blockchain network. They allow for automatic execution of transactions, while encapsulating information. They provide an interface for control access to the ledger and allow complex business logic to be embedded in the network. Hyperledger Fabric smart contracts are called chaincode. When external applications wish to interact with the blockchain, the chaincode for the associated application is invoked. Developers can use Go, Javascript or Java for the development of chaincode. Execution of the chaincode is initiated through a transaction proposal, and is executed against the ledger’s state database at that time instant. Chaincode invocations result in the creation of state transitions. 

\subsubsection{Channels}
An immutable ledger and chaincode is assigned on a per-channel basis. The deployed chaincode can change and manipulate the state of the ledger. The channel dictates the scope of the ledger. If every participant in the network is a part of the channel, then a common shared ledger is available to all the participants. Channels can also be privatised to include only a specific set of participants, thereby allowing the participants to segregate their transactions and ledger. 

\subsubsection{Peers}
Peers are the nodes in the network. They form the fundamental element within the network. They host the ledgers and the chaincode written for the network. There are different types of peers in a hyperledger fabric network, namely:
\begin{itemize}
\item Endorser Peers:
They simulate and endorse the transaction requested by the client.

\item Commiting Peers:
Their job is to verify the transactions and create a consensus and add it to the blockchain.

\item Client Peers:
Using the fabric Software Development Kit, REST servers can be created with which the user applications can be developed to interact with the network.
\end{itemize}

\subsubsection{Orderers}
Since hyperledger fabric is permissioned blockchain and follows a deterministic consensus mechanism, rather using probabilistic consensus mechanisms like the one used by Bitcoin\cite{nakamoto2019bitcoin} or
Ethereum\cite{buterin2014ethereum}, the blockchain cannot be forked into different versions which would lead to inconsistency in the ledger. Thus an orderer node is used. The orderer creates a final and corrected block verified by all the peers and distributed to all peer nodes using a messaging service like Kafka or RabbitMQ.

\subsubsection{Membership Service Provider}
Since hyperledger fabric is a permissioned blockchain, identities are provided to the users. A Public Key Infrastructure is used for creating these identities. A user has a public and private key. The user has to sign transactions with the private key, the MSP on the orderer then verifies it with the users public key. The private key is used to produce a signature on a transaction that only the corresponding public key, that is part of an MSP, can match. Thus, the MSP is the mechanism that allows that identity to be trusted and recognized by the rest of the network without ever revealing the member’s private key.

\subsection{Interplanetary File System}
Blockchain technology is not suitable for storing files of massive sizes. It is an expensive medium for storage of files of medium to large sizes. Electronic medical records such as MRI’s generate files of sizes often upwards of 200 MegaBytes. This makes using the blockchain for storage of these files infeasible due to its inherently high network latency. The Interplanetary File System (IPFS) is a protocol which provides decentralized and distributed file storage solution that identifies each file content stored using Content Identifiers (CIDs) . Every file on the IPFS has a unique CID or SHA-256 hash value associated with it, which makes the sharing of the file straight forward regardless of the size of their underlying content. No information, other than the associated hash value is required to access the file. The file is duplicated across multiple storage nodes depending upon the frequency of access. Due to this duplication, IPFS supports a high level of concurrent access and throughput. The hash size of the files stored on the IPFS is also only a couple of tens of bytes, which makes the storage of the hash values of the associated IPFS files on the blockchain viable. The hash values of the medical files uploaded to the IPFS would be stored in the Fabric Ledger. 

\section{Proposed Platform}
Using Hyperledger Fabric and the IPFS, the authors of this paper propose a decentralised medical data sharing platform. Current healthcare solutions dissolve the patient of the right to control their own digital healthcare identity. Moreover it makes it extremely difficult for the patient to share their medical data because of the heterogeneity in the healthcare solutions adopted by medical institutions. 

Users of the system will register with the Certification Authority. Multiple Certification Authorities, managed by the different hospitals would be a part of the Fabric network. The Certification Authority provides X509 certificates to all the participating components in the network. These certificates are required by the components of the system to access the client nodes.
Once registered, users of the system would be able to run chaincode which dictates the access control list and the implementation logic. The chaincode is deployed on the peer nodes in the network of all the hospitals. This chaincode prior to deployment has to be approved by all the participating hospitals. 
On registration, every hospital is given a unique public-private key pair. This cryptographic material are used for encryption of the data generated at the hospital, as well as to digitally sign it. Data that is encrypted with the private key, can only be decrypted with the public key. The private key of the hospital is kept secret. The public key can be shared with other healthcare institutions, using it to decrypt the data shared, thus verifying the provenance of the medical file.
A Hyperledger Fabric channel is a private “subnet” of communication between two or more specific network members, for the purpose of conducting private and confidential transactions. A channel is defined by members (organizations), anchor peers per member, the shared ledger, chaincode application(s) and the ordering service node(s). Each transaction on the network is executed on a channel, where each party must be authenticated and authorized to transact on that channel. Each peer that joins a channel, has its own identity given by a membership services provider (MSP), which authenticates each peer to its channel peers and services. In this paper the authors have only implemented a single channel between all the peers, shared ledger, chaincode application and ordering nodes.

\begin{figure}[ht]
\centering
\includegraphics[scale=0.3]{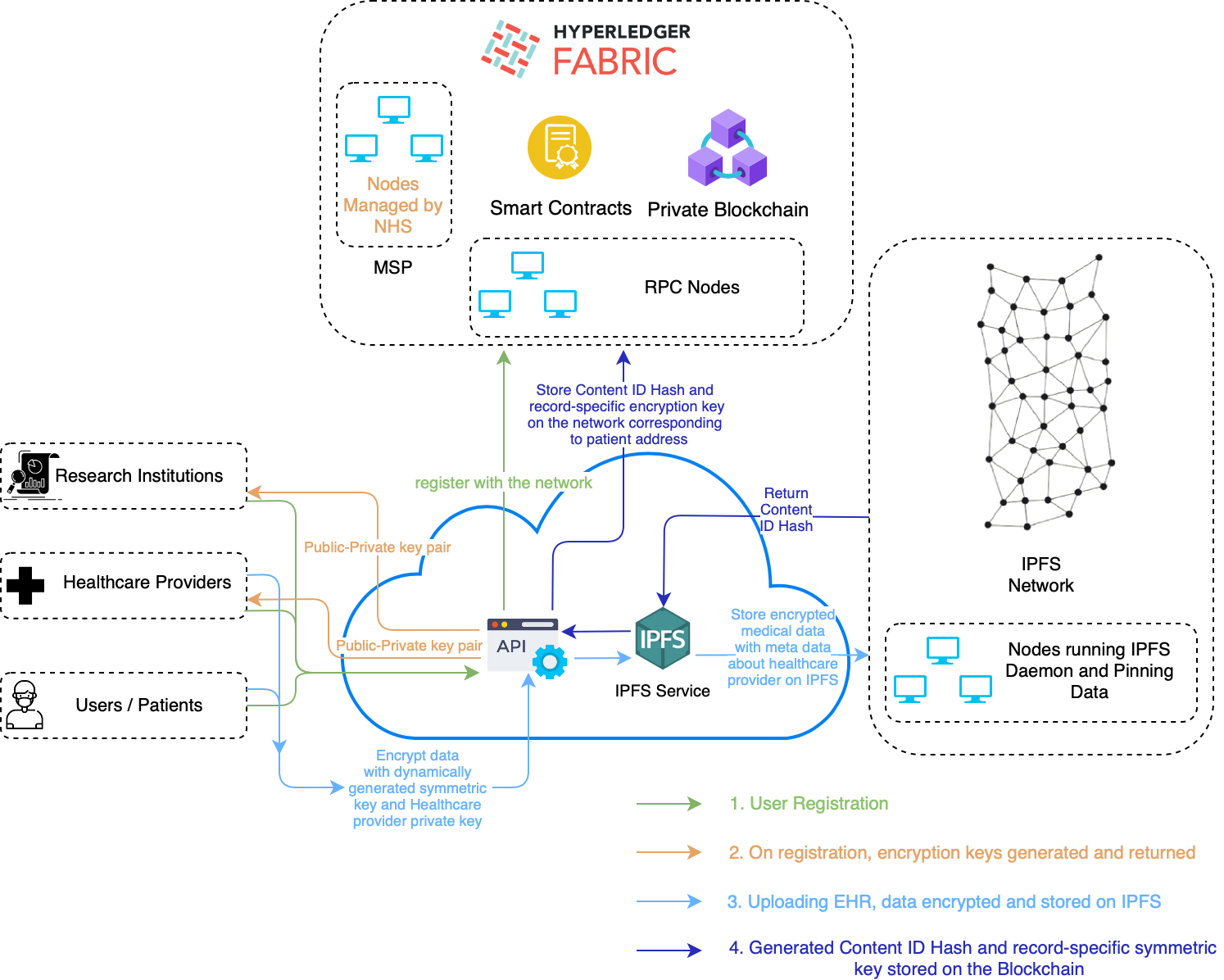}
\caption{User registration and uploading patient EHRs}
\label{fig:overview}
\end{figure}

When a patient wishes to add their medical files to the network, a record-specific symmetric key is dynamically generated. The private key of the healthcare provider who created the medical record and the record-specific symmetric key of the patient is used to encrypt the file. This file, along with information about the healthcare provider is bundled and stored on the IPFS. The IPFS uses a content addressing system, which uses Content Identifiers (CID) to locate specific data. For the bundled data uploaded to the IPFS, a unique CID is returned. This CID is stored on the Blockchain network along with the record-specific symmetric key corresponding to the patient's address. Any access to any medical record would require decryption using the Healthcare Institutions public key, and the patients symmetric key.

\begin{figure}[H]
\centering
\includegraphics[scale=0.15]{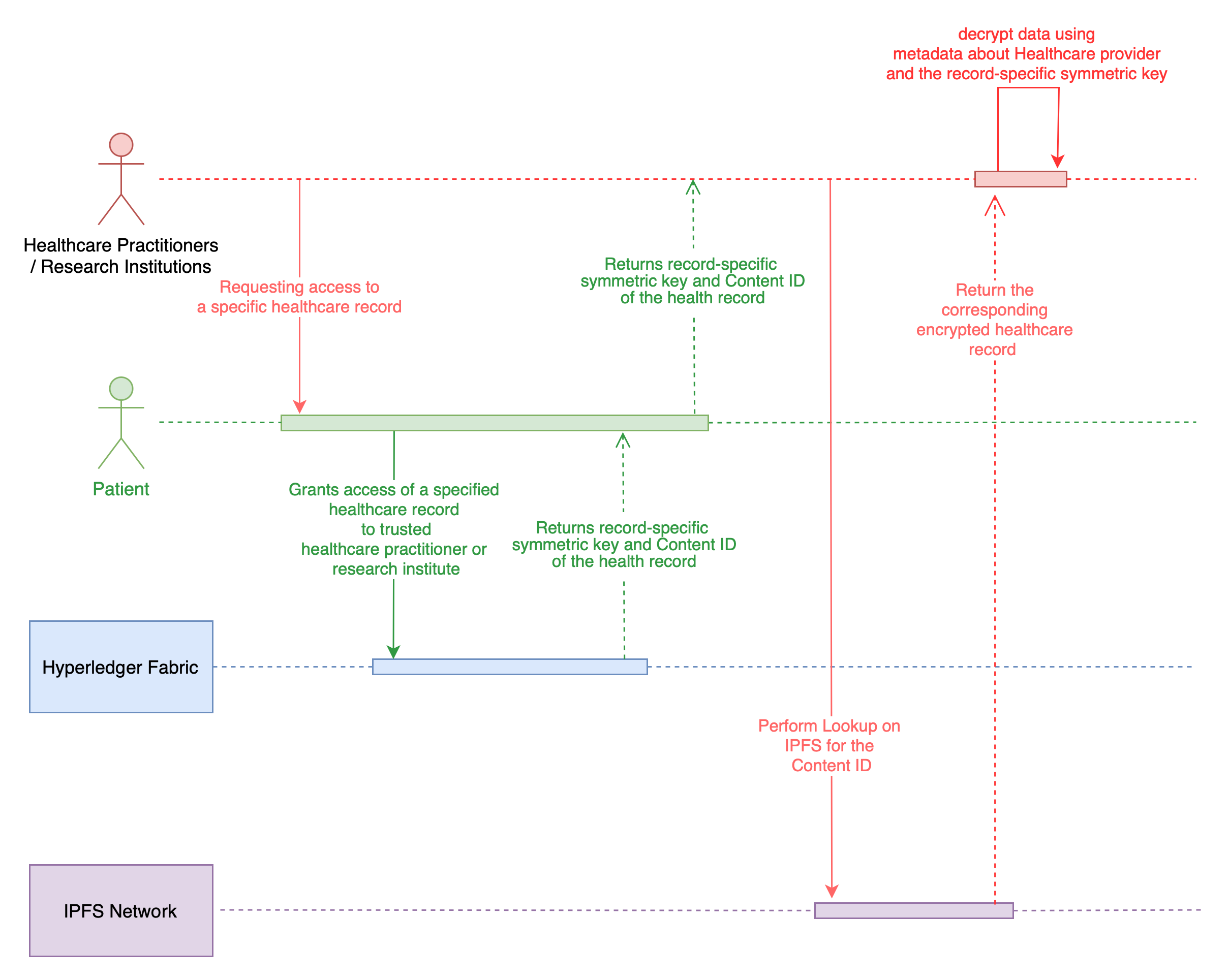}
\caption{Sharing a single medical file with Healthcare Practitioners and Research Institutions}
\label{fig:shareWithDoc}
\end{figure}

When the patient wishes to share a specific medical record with a healthcare practitioner or a research institution, they send a request to the client node to allow access to the medical record. The record-specific symmetric key and the CID for the encrypted file on the IPFS is returned as a result of the request. The patient shares the symmetric key and the CID with the healthcare practitioner or research institution. Using the above mentioned cryptographic materials, the healthcare practitioner or the research institution can decrypt the medical record. The decryption happens on the local machines of those entities with whom the cryptographic material is shared. Sophisticated logic is embedded in the chaincode of the channel used for sharing the data, which automatically transfers tokens to the patients account.

 In the proposed platform, healthcare institutions run a IPFS pinning service. The pinning service handles the storage of the patient data on physical nodes. Healthcare institutions in the system manage nodes that run the IPFS pinning service. The management of the nodes is specific to the healthcare institution, but the data pinned on the nodes is accessible to anyone with the CID of the data stored on it. The patients themselves can pin their data on their personal machines with a IPFS service running, provided they know their data's CID. Patient data would be pinned across multiple nodes. Thus, the patient's data is hosted on multiple locations instead of a single private location. Data stored on the IPFS is addressed by a unique CID generated specific to that data. Any change in the data would lead to a completely different CID being generated. Thus, the CID stored on the blockchain is guaranteed to point to the data that was uploaded by the healthcare institution and patient. 

\section{Problems Addressed}

\begin{itemize}
\item Data Access Management \\
An implicit property of data stored in a blockchain is attribution. Ownership of data can be managed by sophisticated logic dictated by powerful chaincode. This allows for variation in the granularity of data accessed, allowing for parts of data to be available while the remaining is inaccessible. Hyperledger Fabric's chaincode can enable patients to take control of their own data, and manage their virtual identity. Patients can control who can read their medical history or data, specify the time limit for such access and monetize data sharing by receiving digital or physical assets in exchange for access rights. It could change the data ownership landscape by placing control of the patient’s data with patients themselves, and not with organisations that create and manage it. 

\item Data Availability \\
Due to the immutable nature of blockchains, data stored on the chain is guaranteed to maintain integrity. On-chain medical records or links to the off-chain storage for these records stored on the blockchain are thus safeguarded from tampering. The distributed nature of blockchain also enables high availability. Using the IPFS for storing the medical data ensures a highly robust system. Distributing encrypted data across multiple locations ensures that access to it is not hindered and has no singular point of failure. Encapsulating the patient from the implementation logic, it would also allow the patient to access their medical records in one location, instead of sifting through enterprise specific data storage which is difficult to access and share.

\item Data Acquisition \\
Gathering reliable and ethically sourced healthcare data continues to be a contentious issue for the medical industry. With data ownership belonging to patients, and with access rights managed by them, pharmaceutical companies and research institutions can request access to specific data in exchange for digital or physical assets. Controlling which aspects of their medical records they want to share; patients can maintain a level of control that they previously would not be able to. Not only does this incentives data sharing, but also ensures the integrity of data that the industry has access to. This potential access to large amounts of medical records, the veracity of which is guaranteed by the blockchain, opens doors to many possibilities for the industry. Analysing the medical records and finding correlations between occurrences of certain conditions, the pharmaceutical industry can create targeted medicine that may cure, halt or prevent diseases more effectively.

\item Data Sharing \\
Although the proposed solution may not be able to solve the issue of standardizing all the data formats used throughout the industry, it provides a common interface through which data can be securely and ethically stored, accessed and utilised. This holds immense value to doctors and patients. Doctors would not have to rely on the patient’s diligence to safely store their medical records or tests to access during treatments, instead they would easily locate it on the blockchain. Patients would also save time and money, as they would not have to redundantly perform the same tests. Using IPFS as the storage solution, and the powerful data access controls that Hyperledger Fabric provides, patients can share their data with concerned individuals. This can also be used to further compensate data sharing and lead to provide tangible assets in exchange.

\item Improved Research \\
A reason why many Artificial Intelligence and machine learning based medical innovations haven’t been able to find relevance in developing or underdeveloped countries is their inability to generalise. Research conducted and the models that are created, often done so in developed nations. These models use data from these advanced, state-of-the-art machines. Patients in developing or underdeveloped nations are often limited to older equipment that outputs relatively lower quality and resolution data. Thus, the solutions that are created by researchers, often preclude people from impoverished nations. Medical conditions that are prevalent in people from these nations, may not be so in the data sets  used to create the solution, thus are not representative of those people.
The proposed platform has the potential to create a global system for medical data sharing. This would enable researchers to obtain diverse data sets that resemble the real world. They could procure data from populations peripheral in current medical data sets, and create solutions that would be widely applicable to a broader audience. 

\item Personalised Medicine \\
Better research would result due to access to reliable medical data. This opens up the possibility to conduct data analysis of large medical data sets and find patterns within then. As the platform would provide a highly secure medium for data exchange, it can be used for the sharing of genomic data. This data can be shared with pharmaceutical companies for research and development while being compensated for doing so. Thus, it would empower the industry to create personalised medicine, using the reliable data that they would have access to.

\end{itemize}

\bibliography{bibliography}

\begin{thebibliography}{1}
\urlstyle{rm}
\expandafter\ifx\csname url\endcsname\relax
  \def\url#1{\texttt{#1}}\fi
\expandafter\ifx\csname urlprefix\endcsname\relax\def\urlprefix{URL }\fi
\expandafter\ifx\csname doiprefix\endcsname\relax\def\doiprefix{DOI: }\fi
\providecommand{\bibinfo}[2]{#2}
\providecommand{\eprint}[2][]{\url{#2}}

\bibitem{leeming2019ledger}
\bibinfo{author}{Leeming, G.}, \bibinfo{author}{Cunningham, J.} \&
  \bibinfo{author}{Ainsworth, J.}
\newblock \bibinfo{journal}{\bibinfo{title}{A ledger of me: personalizing
  healthcare using blockchain technology}}.
\newblock {\emph{\JournalTitle{Frontiers in medicine}}}
  \textbf{\bibinfo{volume}{6}} (\bibinfo{year}{2019}).

\bibitem{team2018medibloc}
\bibinfo{author}{Team, M.}
\newblock \bibinfo{title}{Medibloc whitepaper, 2017} (\bibinfo{year}{2018}).

\bibitem{thwin2018blockchain}
\bibinfo{author}{Thwin, T.~T.} \& \bibinfo{author}{Vasupongayya, S.}
\newblock \bibinfo{title}{Blockchain based secret-data sharing model for
  personal health record system}.
\newblock In \emph{\bibinfo{booktitle}{2018 5th International Conference on
  Advanced Informatics: Concept Theory and Applications (ICAICTA)}},
  \bibinfo{pages}{196--201} (\bibinfo{organization}{IEEE},
  \bibinfo{year}{2018}).

\bibitem{nizamuddin2018ipfs}
\bibinfo{author}{Nizamuddin, N.}, \bibinfo{author}{Hasan, H.~R.} \&
  \bibinfo{author}{Salah, K.}
\newblock \bibinfo{title}{Ipfs-blockchain-based authenticity of online
  publications}.
\newblock In \emph{\bibinfo{booktitle}{International Conference on
  Blockchain}}, \bibinfo{pages}{199--212} (\bibinfo{organization}{Springer},
  \bibinfo{year}{2018}).

\bibitem{nakamoto2019bitcoin}
\bibinfo{author}{Nakamoto, S.}
\newblock \bibinfo{title}{Bitcoin: A peer-to-peer electronic cash system}.
\newblock \bibinfo{type}{Tech. Rep.}, \bibinfo{institution}{Manubot}
  (\bibinfo{year}{2019}).

\bibitem{buterin2014ethereum}
\bibinfo{author}{Buterin, V.} \emph{et~al.}
\newblock \bibinfo{journal}{\bibinfo{title}{Ethereum: A next-generation smart
  contract and decentralized application platform}}.
\newblock {\emph{\JournalTitle{URL
  https://github.com/ethereum/wiki/wiki/5BEnglish5D-White-Paper}}}
  \textbf{\bibinfo{volume}{7}} (\bibinfo{year}{2014}).

\bibitem{androulaki2018hyperledger}
\bibinfo{author}{Androulaki, E.} \emph{et~al.}
\newblock \bibinfo{title}{Hyperledger fabric: a distributed operating system
  for permissioned blockchains}.
\newblock In \emph{\bibinfo{booktitle}{Proceedings of the thirteenth EuroSys
  conference}}, \bibinfo{pages}{1--15} (\bibinfo{year}{2018}).

\end{thebibliography}

\end{document}